# Quantum mechanics can provide unbiased result

**Arindam Mitra**

Anushakti Abasan, Uttar Phalguni -7, 1/AF, Salt Lake,
Kolkata, West Bengal, 700064, India.

**Abstract:** Getting an unbiased result is a remarkably long standing problem of collective observation/measurement. It is pointed out that quantum coin tossing can generate unbiased result defeating dishonesty.

Honesty is said to be the best policy. The moral can be scientifically proved to be right if honest person can scientifically defeat dishonest person in any case or if we are scientifically compelled to be honest to accomplish a task. So far nothing – no science, no technology, no religion, no moral lesson etc - compels us to be honest to solve any problem.

There is no scientific basis of the word " unbiased". It is still a concept. So far two people cannot generate an unbiased result in any observation or measurement. This problem has been studied in two-party coin tossing [1-7]. This is not merely a problem of collective observation/measurement. It reflects our endless fight against dishonesty.

While addressing this problem we have observed [8] that Alice can prevent Bob from being dishonest, but Bob cannot prevent Alice. Therefore, Bob cannot bias the outcome of tossing but Alice can. Here we shall see that both Alice and Bob can prevent each other from being dishonest. Therefore, Alice and Bob will be scientifically compelled to generate unbiased result honestly.

It is known that unbiased outcome can be easily generated if coin tossing algorithm is build on cheating-free bit commitment algorithm. But no cheating-free bit commitment algorithm has been found so far. In bit commitment, Alice commits a bit towards Bob without revealing the bit value and later Alice unveils the bit value by disclosing some information. Here the concern is, Alice can cheat Bob by changing the bit value and Bob can cheat Alice by recovering the bit value before its disclosure.

With the advent of quantum information it was thought [9] that quantum bit commitment (QBC) algorithm will be cheating-free. However, in 1997 Mayers, Lo and Chau claimed [10,11] that no QBC algorithm will be cheating-free due to quantum-computer-assisted entanglement attack. It was also claimed [12,13] that it is impossible to achieve zero bias with quantum coin tossing. But this claim is the logical reverberation of the former claim. This development is quite frustrating [14,15].

In their analysis [10,11] authors basically considered standard coding where qubit - two-state quantum state- is analogous to bit. Therefore, their result is applicable to all QBC and QCT algorithms based on standard coding. But the problem is, without any proof authors claimed that the result would be applicable for any QBC and QCT algorithm. Here authors missed the point that alternative QBC and QCT algorithm may exist. As a result they abruptly claimed that in any QBC algorithm, where density matrices associated with bit 0 and 1 are same, the bit committed can be changed with probability 1. The same mistake has been propagated in the subsequent works. It has also been claimed [16] that in any QBC scheme it is impossible to arbitrarily conceal the bit committed till its disclosure.

After the publication of the papers [10,11] an alternative quantum coding (AQC) technique has been developed [17,18] where a sequence of quantum states represents a bit. We shall see that alternative QCT algorithm can be built on alternative QBC algorithm to generate unbiased outcome. Interestingly, we need entanglement and quantum computer/memory to accomplish this task.

First we shall present a quantum coding on which alternative QBC algorithm will be built. Suppose there are some EPR pairs of spin-1/2 particles. EPR pairs can be arranged either in direct order or in reverse order. Let the direct and reverse order of arrangements of EPR pairs be.

$$S_i^0 = \{A, B, C, D, E, F, G, H\}$$
$$S_j^0 = \{A, B, C, D, E, F, G, H\}$$

$$S_i^1 = \{A, B, C, D, E, F, G, H\}$$
$$S_j^1 = \{H, G, F, E, D, C, B, A\}$$

The same pair of letters denotes an EPR pair. In $S_i^0$ and $S_j^0$, EPR pairs are arranged in direct order. In $S_i^1$ and $S_j^1$, EPR pairs are arranged in reverse order. Two pairs of entangled sequences, representing bit 0 and 1, are made up of the four (1:1:1:1) EPR states written as

$$|\psi_\pm\rangle_{i,j} = \frac{1}{\sqrt{2}}\left(|\uparrow\rangle_i|\downarrow\rangle_j \pm |\downarrow\rangle_i|\uparrow\rangle_j\right)$$

$$|\varphi_\pm\rangle_{i,j} = \frac{1}{\sqrt{2}}\left(|\uparrow\rangle_i|\uparrow\rangle_j \pm |\downarrow\rangle_i|\downarrow\rangle_j\right)$$

where i,j denotes the position of an EPR pair in $S_i$ and $S_j$ and "↑" and "↓" denote two opposite spin directions. Note that these two pairs of arrangements are maximally distinguishable because there is no-overlapping in the arrangements. Note that here density matrices associated with bit 0 and 1 are same; $\rho^0 = \rho^1 = \frac{1}{4}\mathbf{I}$.

Our alternative QBC algorithm can not only be used to commit random bits but also to commit message directly. This is a consequence of AQC. The alternative QBC algorithm is presented below.

**1.** Alice collects N singlets where N need not to be large number. She stores the singlets in her quantum memory and chooses $n_a$ pairs at random

**2**. Alice measures the spin component of each chosen pair along either a fixed secret axis or randomly chosen axis. If Alice gets 100% anti-correlated data (singlet is rotationally symmetric) she proceeds for the next step.

**3.** Alice applies unitary operators $U^a \in \{\sigma_x, \sigma_y, \sigma_z, I\}$ at random on each of remaining particles $A_k$ where $\sigma_x, \sigma_y, \sigma_z$ are Pauli matrices and I is $2 \times 2$ identity matrix. The shared ensemble can be described [8] by $\rho = \frac{1}{4}I$ where I is $4 \times 4$ identity matrix.

**4**. Alice transmits the particle $B_k$ of each pair $A_k B_k$ to Bob.

**5.** To verify shared entanglement [8] Bob requests Alice to send $n_b$ particles $A_k$ after selecting them at random.

**6**. Alice applies the appropriate $U^a$ on her particle to convert the chosen pairs into singlets and sends the her particles $A_k$ to Bob.

**7**. Bob measures spin components of $n_b$ pairs $A_k B_k$ along either a fixed axis or randomly chosen axes. If Bob gets 100% anti-correlated data he proceeds for the next step.

**8**. Bob applies unitary operators $U^b \in \{\sigma_x, \sigma_y, \sigma_z, I\}$ at random on the remaining particles $B_k$.

**9.** Alice measures the spin component of each of the remaining n shared pairs along x and z axis at random.

**10.** To commit bit 0, Alice reveals results, (d) in direct order ($d_k$). To commit bit 1, Alice reveals results in reverse order ($d_{n-k}$) Revealing results in the direct and reverse order is tantamount to arranging particles in direct and reverse order respectively.

**11.** To reveal bit value Alice discloses the information of her $U^a$ and chosen axes.

**12.** From the available information regarding $U^a$ and $U^b$ Bob identifies the positions of the singlets in their final shared sequence-ensemble. He then measures the spin components of his singlet particles along Alice's chosen axes. There is no need to measure the spin component of his other EPR particles. In their data, if Bob gets 100% anti-correlation in direct order ($d_k d_k$) and 50% anti-correlation in reverse order ($d_k d_{n-k}$) the bit committed is 0. But if Bob gets 100% anti-correlation in reverse order and 50% anti-correlation in direct order, the bit committed is 1.

Results can be revealed by two symbols. Suppose, "up" and "right" spin directions are denoted as *1*, and "down and "left" spin direction are denoted as *0*. The results itself do not reveal any state information because $|\uparrow\rangle = \frac{1}{\sqrt{2}}(|\rightarrow\rangle + |\leftarrow\rangle)$ and $|\downarrow\rangle = \frac{1}{\sqrt{2}}(|\rightarrow\rangle - |\leftarrow\rangle)$.

It is easy to see that it is impossible for Alice to change the bit committed and for Bob to know the bit committed before the unveiling step.

***Proof:*** Density matrices of the equal mixture of four EPR-Bell states and equal mixture of the direct product states $|\uparrow\uparrow\rangle, |\uparrow\downarrow\rangle, |\downarrow\uparrow\rangle$ and $|\downarrow\downarrow\rangle$ are same.

$$\rho = \frac{1}{4}(|\psi_-\rangle\langle\psi_-| + |\psi_+\rangle\langle\psi_+| + |\varphi_-\rangle\langle\varphi_-| + |\varphi_+\rangle\langle\varphi_+|) = \frac{1}{4}\mathbf{I}$$

$$\rho = \frac{1}{4}(|\uparrow\downarrow\rangle\langle\uparrow\downarrow| + |\downarrow\uparrow\rangle\langle\downarrow\uparrow| + |\uparrow\uparrow\rangle\langle\uparrow\uparrow| + |\downarrow\downarrow\rangle\langle\downarrow\downarrow|) = \frac{1}{4}\mathbf{I}$$

where I is $4\times 4$ identity matrix. These two equations imply that EPR correlation would be completely suppressed in $\rho = \frac{1}{4}\mathbf{I}$. Here EPR correlation is independently suppressed by Alice and Bob.

These two equations imply that Bob will get totally uncorrelated data until and unless Alice discloses how she prepared $\rho = \frac{1}{4}\mathbf{I}$ It is impossible for Bob to recover the bit committed from uncorrelated data.

These two equations imply that Bob's data will be uncontrollable random data and it will remain unknown to Alice because she never knows how Bob prepared $\rho = \frac{1}{4}\mathbf{I}$. As Bob's data are uncontrollable and unknown to Alice, it is impossible for her to reverse the order of EPR correlation by any operation or trick. Therefore, it is impossible for Alice to change the bit committed.

The proof will break down if any of the two parties does not suppress entanglement. Suppose they all along share singlets. Then Bob can know bit value before unveiling step. Bob can measure spin component choosing x and z axis at random. He will get 75% anti-correlation which is enough for recovery of bit value. Suppose Bob is not interested to know before its disclosure. Then Alice can reverse the commitment. To reverse the commitment Alice has to identify 25% cases ($d_k d_{n-k}$) where her own data corresponding to the k-th position will be perfectly anti-correlated with Bob's data corresponding to (n-k)-th position. Now Alice can safely claim that in those positions states are singlets.

Probabilistically, it is possible to *evade* fundamental impossibilities. As for example, a bit of information can be communicated faster than light with probability 1/2. Note that non-EPR state and measurement of spin components of an EPR pair along non-identical axis can generate correlated data with optimal probability 1/2. Therefore, even if Alice does not use EPR states Bob can get correlated data either in direct order or in reverse order with optimal probability $\frac{1}{2^n}$ where *n* is Bob's chosen statistics. By measurement Bob know the bit committed before Alice's disclosure with optimal probability $\frac{1}{2^n}$. This is not the Bob's optimal probability of knowing the bit before Alice's disclosure. By guessing Bob can know the bit with probability 1/2. Similarly, by revealing wrong information about her chosen axis and $U^a$ Alice can reverse the order of correlation with optimal probability $\frac{1}{2^n}$ where *n* is the chosen statistics. By revealing wrong information

Alice can change the bit committed with optimal probability $\frac{1}{2^n}$. This is also not Alice's optimal probability of changing the bit committed. Without revealing the bit value Alice can change the bit committed with optimal probability 1/2 because it that case Bob has to guess the bit value.

Coin tossing scheme can be built on bit commitment scheme in the following way. i) Alice commits a bit. ii). Bob guesses the bit and reveals his guess-bit to Alice. ii). Alice reveals the bit value which she committed. If Bob's guess is right, the protocol would generate a bit, say 1. Otherwise 0.

Alternative QCT algorithm can be built on our alternative QBC algorithm as described above. Bob's probability of generating a preferred bit is $p_B = \frac{1}{2} + \varepsilon_B$ where $\varepsilon_B$ is the bias given by Bob. Bob's probability of generating a preferred bit will be equal to Bob's optimal probability of knowing the bit value before its disclosure. So, $p_B = \frac{1}{2}$ and $\varepsilon_B = 0$. Alice's probability of generating a preferred bit is $p_A = \frac{1}{2} + \varepsilon_A$ where $\varepsilon_A$ is the bias given by Alice. Alice's probability of generating a preferred bit will be equal to Alice's optimal probability of changing the bit committed. So, $p_A = \frac{1}{2}$ and $\varepsilon_A = 0$.

We did not consider noise, but noise should be considered because one can try to manipulate noise to cheat other. Due to noise nothing can be perfect. Neither Alice nor Bob can prepare the state $\rho = \frac{1}{4}I$ with imperfect unitary machine. But here the advantage is, $\rho = \frac{1}{4}I$ can be prepared by simply introducing noise. Before sharing EPR particles firstly Alice and Bob have to determine environmental noise levels in the singlets, say $L_a$ and $L_b$. Secondly, Alice and Bob have to determine the noise levels, $L_A$ and $L_B$, required to suppress entanglement after applying $U^a$ and $U^b$. So before transmitting particles Alice introduces $L_A$, and before Alice's commitment Bob introduces $L_B$. Still the problem is, one can use better detector than others' detector to unlock entanglement. As detector noise cannot be separately determined Alice and Bob separately additionally introduces $L_a$ and $L_b$ respectively to prevent unlocking of suppressed entanglement. Thus noise can be used to prevent the manipulation of noise.

Following the above mentioned technique entanglement can be doubly suppressed by Alice and Bob in presence of noise. This will be suppressed due to environmental noise, unitary operations and self introduced noise. But self introduced noise will corrupt more singlets. But as long as noise level for the singlets is within the tolerable limit Bob will be able to recover bit value with high statistical confidence level. Let us take $N = 300$, $n_a = 50$ and $n_b = 50$. Then, $n = 200$. The bit will be encoded on approximately 50 singlets. Now if we consider 25% over all noise, still the bit value can be recovered with high statistical confidence level.

Coin tossing may not be an important issue. But this relatively unimportant issue significantly reveals that quantum mechanics can absolutely prohibit dishonesty. Quantum mechanics may be regarded as a discipline of moral science.

**Note added:** This work and some other works [19,20] possibly give a hint that quantum mechanics provides unparallel protection not against eavesdropping, but against cheating.

*email:mitra1in@yahoo.com